\documentclass[fleqn,twoside]{article}

\usepackage[headings]{espcrc2}
\usepackage{graphicx}
\usepackage{epsfig}
\usepackage[figuresright]{rotating}
\usepackage{amsmath}

\newcommand{\be}{\begin{equation}}
\newcommand{\ee}{\end{equation}}
\newcommand{\bea}{\begin{eqnarray}}
\newcommand{\eea}{\end{eqnarray}}

\def \litwo   {{\rm{Li_2}}}

\title{
\vspace*{-2.5cm}
\begin{flushleft}
\ttfamily{DESY 08-084 \\
PITHA-08/14}
\end{flushleft}
Virtual Hadronic Corrections to Massive Bhabha Scattering}

\author{
Stefano~Actis%
\address[AC]{Institut f\"ur Theoretische Physik E, RWTH Aachen University, D-52056 Aachen, Germany}, 
Janusz~Gluza%
\address[KA]{Institute of Physics, University of Silesia, Uniwersytecka 4, PL-40007 Katowice, Poland},
Tord~Riemann%
\address[ZE]{Deutsches Elektronen-Synchrotron, DESY, Platanenallee 6, D-15738 Zeuthen, Germany}
}

\runtitle{}
\runauthor{}

\begin{document}

\begin{abstract}
Virtual hadronic contributions to the Bhabha process at the NNLO level are discussed.
They  are substantial for predictions with per mil accuracy.
The  studies of heavy fermion and hadron corrections complete the calculation of Bhabha virtual effects at the NNLO level.
\end{abstract}

\maketitle

\section{Introduction}
Since Loops and Legs 2006 
\cite{Blumlein:2006rv,Actis:2006dj}, 
considerable progress in the determination of the virtual NNLO corrections 
to the massive QED Bhabha process has been made \footnote{For earlier literature, see 
\cite{Arbuzov:1995vj,Arbuzov:1998du,Bern:2000ie,Glover:2001ev,Bonciani:2003cj,Czakon:2004tg,Czakon:2004wm,Bonciani:2004gi,Bonciani:2004qt,Penin:2005kf,Penin:2005eh,Bonciani:2005im}.}.

The $N_f=2$ two-loop corrections with heavy-fermion insertions have been computed
in the limit $m_e^2<<m_f^2<<s,t$: with a direct Feynman diagram calculation in \cite{Actis:2007gi}, and using  
a factorization formula that relates massless and massive amplitudes in \cite{Becher:2007cu}.

Hadronic contributions have been recently calculated in \cite{Actis:2007fs}.
In addition, heavy-fermion corrections, beyond the  $m_f^2<<s,t$ limit,
have been made available in \cite{Actis:2007fs,Bonciani:2008zz,Bonciani:2008ep}; see also \cite{rbonciani}.
 
Interestingly, the original expectations on the necessity of a complete, direct two-loop massive Feynman diagram evaluation have not been fulfilled yet.
After the analytical evaluation of a massive planar and a massive non-planar double-box diagram (both with seven propagators) in \cite{Smirnov:1999gc} and in \cite{Tausk:1999vh}, there was hope to evaluate all the remaining diagrams soon \cite{Heinrich:2004iq}. So far only part of the photonic master integrals, namely the planar ones, have been evaluated in \cite{Czakon:2006pa} in the limit of small electron mass.
Although this limit is by far sufficient for experiments, there is still space
for theoretical developments.

Finally, it should be stressed that results for three gauge-invariant classes of NNLO Feynman diagrams 
have been determined by at least two independent groups, relying on different methods:
\begin{itemize}
 \item \emph{Photonic corrections:} computed in \cite{Penin:2005kf,Penin:2005eh} and
recalculated in \cite{Becher:2007cu};
\item \emph{Electron $N_f=1$ corrections:} computed in \cite{Bonciani:2004gi} and cross-checked
in \cite{Actis:2007gi} (with full $m_e$ dependence) and  in \cite{Becher:2007cu} (small electron mass limit);
\item \emph{Heavy-fermion $N_f=2$ contributions:} determined with two independent 
methods in the limit $m_f^2<<s,t$ in \cite{Actis:2007gi,Becher:2007cu} and 
for any mass $m_f$  in \cite{Actis:2007pn,Actis:2007fs} (dispersive approach) and
in \cite{Bonciani:2008zz,Bonciani:2008ep} (analytical result).
\end{itemize}

As far as the full electroweak model is concerned, the
NLO corrections are well  known \cite{Consoli:1979xw,Bohm:1988fg}. 
At NNLO, the terms enhanced by the heavy top quark are taken into account   \cite{Bardin:1990xe,Bardin:1999yd}. 
The full dependence of weak corrections on the top quark and Higgs boson masses is known \cite{Awramik:2003rn,Awramik:2006uz} and  implemented for fermion pair production in $e^+e^-$-annihilation \cite{Arbuzov:2005ma},  but is not yet implemented for Bhabha scattering.
Here, we report on another class of NNLO corrections, which have recently determined:
\begin{itemize}
 \item  \emph{Virtual hadronic NNLO contributions}, including both reducible self-energy insertions and irreducible vertex and box corrections \cite{Actis:2007fs}.
\end{itemize}

For a review on the status of Monte Carlo studies for Bhabha scattering we refer to \cite{gmontagna}. 

\section{Hadronic Virtual Corrections}

\begin{figure}[hb]
\hfill
\includegraphics[scale=0.45]{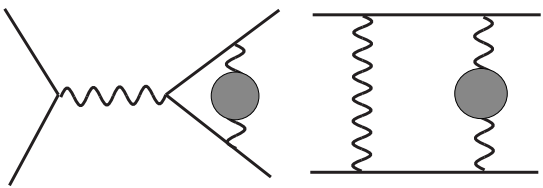}\hfill
\caption{\em
 Two-loop irreducible vertex and box topologies for Bhabha scattering with vacuum
polarization insertions.}
\label{fig1}
\end{figure}

Higher order hadronic corrections to the Bhabha scattering cross 
section can be obtained inserting the renormalized irreducible photon
vacuum polarization function, $\Pi$, in the appropriate virtual-photon propagator 
(see Fig.~\ref{fig1}),
\begin{equation}
\label{1stReplace}
\frac{g_{\mu\nu}}{q^2\!+\!i \delta}  \!\to\!
\frac{g_{\mu\alpha}}{q^2\!+\!i \delta}
\left( q^2 g^{\alpha\beta} \!-\! q^\alpha q^\beta \right) \Pi(q^2)
\frac{g_{\beta\nu}}{q^2\!+\!i \delta},
\end{equation}
where $q$ is the momentum carried by the virtual photon and
$\delta\to 0_+$.
The vacuum polarization function $\Pi$ can be represented by the 
once-subtracted dispersion integral \cite{Cabibbo:1961sz},
\begin{equation}
\label{DispInt}
\Pi(q^2) =
- \frac{q^2}{\pi} \, 
  \int_{4 M^2}^{\infty} \, d z \, 
  \frac{\text{Im} \, \Pi(z)}{z} \, 
  \frac{1}{q^2-z+i\, \delta},
\end{equation}
where the production threshold for the intermediate state in $\Pi$ is located at $q^2=4M^2$.
We leave as understood the subtraction at $q^2=0$ for the renormalized photon
self-energy.

Light-quark contributions get 
modified by low-energy strong-interaction effects, which cannot be computed using
perturbative QCD. 
However, these contributions can be evaluated
using the optical theorem~\cite{Cutkosky:1960sp},
and relating $\text{Im} \, \Pi_{\rm had}$ to the 
hadronic cross-section ratio $R_{\rm had}$~\cite{Cabibbo:1961sz},
\begin{eqnarray}
\label{Rhad0}
\text{Im} \, \Pi_{\rm had}(z)&=& 
- \frac{\alpha}{3} \, R_{\rm had}(z),
\\
R_{\rm had}(z) &=& 
\frac{\sigma(\{e^+e^-\to\gamma^\star\to \text{hadrons}\};z)}
     {(4 \pi \alpha^2)\slash (3z)}. \nonumber \\
\label{Rhad}
\end{eqnarray}
$\text{Im}~\Pi_{\rm had}$ can be computed employing the experimental data for $R_{\rm had}$
in the low-energy region and around hadronic resonances, and the perturbative-QCD
prediction in the remaining regions.

Here we give numerical results  through tables which have been included also 
in the first version of \cite{prlv1}\footnote{In order to
make \cite{Actis:2007fs} less technical, we have replaced in the final version
tables by figures. 
This means that the numbers shown here correspond to the numbers which can be read out 
from \cite{Actis:2007fs}; see also \cite{webPage:2007xx}.}.

\begin{figure}[htb]
\includegraphics[scale=0.3]{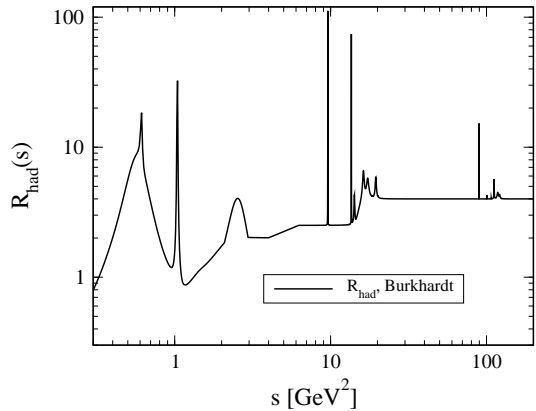}
\caption{\em
 $R_{had}(s)$, based on the parameterization given in \cite{Burkhardt:1981jk}.}
\label{frhad}
\end{figure}

A shape of the $R_{had}$ parameterization we use is given in Fig.~\ref{frhad}.
In a forthcoming publication \cite{hadrPRD}, containing a detailed description of \cite{Actis:2007fs}, 
we will employ a more updated parameterization of $R_{had}$ \cite{Hagiwara:2002ma,Hagiwara:2003da,Hagiwara:2006jt}.
We just mention that the final numbers get modified only slightly 
and do not change qualitatively the situation. 

The lower integration boundary in Eq.~\eqref{DispInt} is given by $M=m_\pi$, where $m_\pi$ is the pion mass.
For self-energy corrections to Bhabha scattering at one-loop order this method was first employed 
in \cite{Berends:1976zn}.

Finally, we note that contributions to $\Pi$ arising from leptons and the top quark can be 
computed directly in perturbation theory, setting $M=m_f$
in Eq.~\eqref{DispInt}, where $m_f$ is the mass of the fermion
appearing in the loop, and inserting the imaginary part of the analytic 
result for $\Pi$.

In the following, we will not consider hadronic effects to the running of the coupling constant;
details can be found in \cite{hadrPRD}. 

\section{Vertex Contributions}
Hadronic irreducible vertex corrections are obtained through the 
interference of the vertex diagrams of Figure~\ref{fig1} with the 
tree-level amplitude. 
Their contribution to the ${\cal O}(\alpha^4)$ differential
cross section is given by: 
\begin{eqnarray}\label{sig-irr-vert}
\frac{ d\sigma_{\rm vert} }{ d\Omega } &=&
4 \left( \frac{\alpha}{\pi} \right)^2 \,
 \left(\frac{  \, \alpha^2 }{2 s}\right) 
   \\
&\times & \Bigl\{ \frac{v_1(s,t)}{s^2}  \text{Re}  V_2(s) 
+\frac{v_1(t,s)}{t^2}    V_2(t) \nonumber \\
&+& \frac{v_2(s,t)}{s,t}  \Bigl[ \text{Re} V_2(s) + V_2(t) \Bigr]
 \Bigl\} + {\cal O}(m_e^2), \nonumber
\end{eqnarray}
where we define $v_1(x,y) =x^2 + 2 y^2 + 2 x y$ and
$v_2(x,y) =(x + y)^2$).
Here $V_2$ summarizes all two-loop fermionic corrections to the QED Dirac form factor,
whose computation can be traced back to the  seminal work of 
\cite{Barbieri:1972as,Barbieri:1972hn}.
The full result can be organized as
\begin{eqnarray}
\label{fullVertex}
V_2(x)&=& V_{2e}(x) + V_{2{\rm rest}}(x),
\end{eqnarray}
where $V_{2e}$ denotes the electron-loop component, see \cite{Burgers:1985qg}.

Heavy-fermion and hadronic contributions, instead, can be evaluated as in 
Ref.~\cite{Kniehl:1988id} through the dispersion integral
\begin{equation}
\label{eq:dispVer}
V_{2{\rm rest}}(x) = \int_{4 M^2}^{\infty} \, dz \, \frac{R(z)}{z} \, K_V(x+i\delta;z),
\end{equation}
where $R$ is given by
\begin{eqnarray}
\label{eq:finalR}
R(z)&=& R_{\rm had}^{(5)}(z) - \frac{3}{\alpha}\,\sum_{f=e,\mu,\tau,t} \,\text{Im}\,\Pi_f(z)
\nonumber\\
    &=& R_{\rm had}^{(5)}(z) +\sum_{f=e,\mu,\tau,t}R_f(z;m_f), 
 \nonumber \\
&& \\
\label{rzmf}
R_f(z;m_f)&=&
 Q_f^2 C_f \left(1+ \frac{2m_f^2}{z}\right)
        \sqrt{1-\frac{4m_f^2}{z}}, \nonumber \\
\end{eqnarray}
and where $C_f$ and $Q_f$ denote the color factor and the electric charge.
The two-loop irreducible vertex kernel function $K_V$, in the limit of a vanishing electron mass, reads as:
\begin{eqnarray}\label{eq:kernelV}
K_V(x;z) &=& \frac{1}{3}  \Bigl\{
 -  \frac{7}{8} 
 -  \frac{z}{2 x} 
 + \Bigl( \frac{3}{4} \!+\! \frac{z}{2 x} \Bigr) \ln\frac{-x}{z}
  \nonumber \\
 &-&  \frac{1}{2}  \Bigl( 1 \!+\! \frac{z}{x} \Bigr)^2 
      \Bigl[ \zeta_2 \!-\! \text{Li}_2 
      \left( 1 \!+\! \frac{x}{z} \right) \Bigr]
\Bigr\}.
\nonumber
\end{eqnarray}
Here $\text{Li}_2$ is the usual dilogarithm and $\zeta_2=\litwo(1) = \pi^2\slash 6$.

In Tables~\ref{table:ver1} and ~\ref{table:ver2}
we show numerical values for the various
components of $V_2(x)$ of Eq.~\eqref{fullVertex} for space-like and time-like values 
of $x$ ($t$ and $s$ channel).
For each fermion flavour, we show the 
           result obtained through the dispersion-based approach (first line) and the one
           coming from the analytical expansion (second line).
           
We can see that the latter numbers approach the former ones in regions where the analytical expansions are expected to become good approximations.
When $m_f^2>s$, the entry is suppressed.

\begin{table*}[ht]
\newcommand{\m}{\hphantom{$-$}}
\newcommand{\cc}[1]{\multicolumn{1}{c}{#1}}
\renewcommand{\tabcolsep}{1pc} 
\renewcommand{\arraystretch}{1} 
\begin{center}
\begin{tabular}{@{}|r|r|r|r|r|}
\hline
$\theta=3^\circ$\, $\vert$ \, $\sqrt{s}$ & 1 GeV & 
10 GeV &  $M_Z$ & 500 GeV \\
\hline 
\hline
$e$ &   -5.880  & -28.47      & -80.91   & -151.0 \\
\hline 
     $\mu$ & {\bf -0.005}   &{\bf -0.20}   &{\bf -2.85}  & {\bf -11.8}  \\
 &  $\times$   & 1.04     & -2.78    &   -11.8  \\
\hline 
    $\tau$ & $<$ {\bf 10}$^{-3}$    & $<$ {\bf 10}$^{-2}$  &  {\bf -0.08}  & {\bf -0.8}  \\
 &  $\times$   & $\times$   & 2.26   & -0.5  \\
\hline 
       $t$ &   &  $<$ {\bf 10}$^{-2}$   &  $<$ {\bf 10}$^{-2}$   &  $<$ {\bf 10}$^{-1}$  \\
 &  $\times$    &  $\times$   &  $\times$    &  $\times$   \\
\hline 
had.    &  {\bf -0.004}       &   {\bf -0.20}      &  {\bf -4.08}       &   {\bf -21.5}      \\
\hline 
\hline
$\theta=90^\circ$\, $\vert$ \, $\sqrt{s}$  & 1 GeV & 10 GeV & $M_Z$  & 500 GeV \\
\hline 
\hline
$e$ &-47.44      &  -122.2     & -246.6     & -386.7 \\
\hline 
     $\mu$ & {\bf -0.74}   & {\bf -7.4}  &{\bf -31.4}    &  {\bf -70.6} \\
 &   -0.36   &  -7.4    &  -31.4  & -70.6    \\
\hline 
    $\tau$ &  {\bf -0.01}   & {\bf -0.4}   & {\bf -4.4}   &  {\bf -16.2} \\
 &  $\times$   & 0.3   & -4.4   & -16.2   \\
\hline 
       $t$ &  &  $<$ {\bf 10}$^{-1}$  & $<$ {\bf 10}$^{-1}$    &  {\bf -0.2 }  \\
 &  $\times$    &  $\times$   &  $\times$    & 1.8    \\
\hline 
had.    &  {\bf -0.87}       &   {\bf -12.5}      &    {\bf -67.6}     &  {\bf -172.2}       \\
\hline 
\end{tabular}
\end{center}
\caption[]{\em Contributions to $V_2$ in the $t$ channel for two values of the scattering angle, 
           $\theta=3^\circ$ and $\theta=90^\circ$, $t=-s\, \sin^2(\theta\slash 2)$.}
\label{table:ver1}
\end{table*}

\section{Box Contributions}
\begin{figure}[htb]
\epsfig{figure=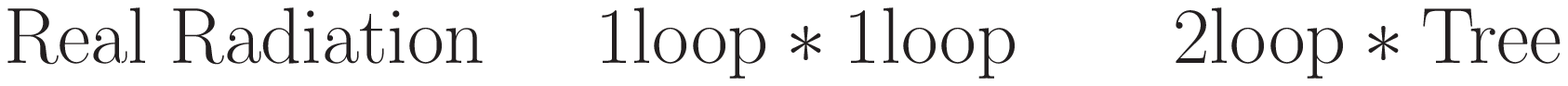,height=0.3cm,angle=90}
\epsfig{figure=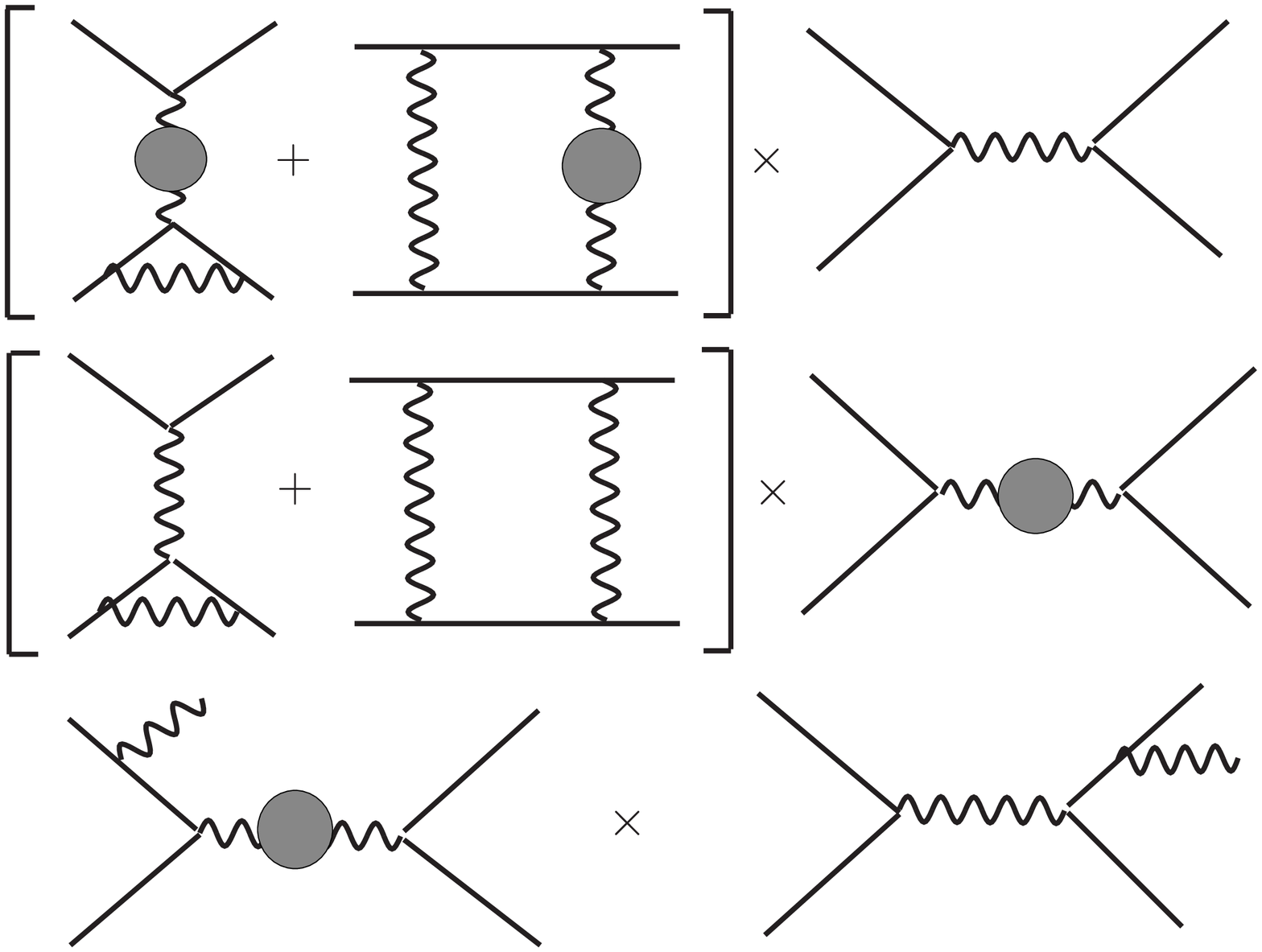,width=7.cm, height=5.5cm}
\caption{IR divergent contributions, becoming finite after combining with irreducible boxes.}
\label{scheme}
\end{figure}

Notice that, unlike the vertex kernel, the irreducible box kernels are infrared divergent, but, 
analogously to the one-loop box diagrams, they have no singularity in the electron mass
\footnote{In \cite{hadrPRD} appropriate simple arguments based on counting of logs will be given.}.
In order to construct an infrared-finite quantity, we combine:
(i) Born diagrams interfering with two-loop box diagrams  and reducible vertices
    (first row in Fig.~\ref{scheme});
(ii) diagrams with a one-loop vacuum polarization insertion interfering with one-loop 
     boxes and vertices (second row in Fig.~\ref{scheme});
 (iii) real single-photon emission diagrams with a one-loop vacuum polarization  insertion 
       (third row  in Fig.~\ref{scheme}).
The infrared-safe $N_f=2$
 irreducible vertices (see Fig.~\ref{fig1}) and pure self-energy diagrams are not included here.

The resulting cross-section  becomes
\begin{eqnarray}
\label{sig1} 
&&\frac{ d\overline{\sigma} }{ d\Omega }=
c \int_{4M_{\pi}^2}^{\infty} dz \frac{R_{\mathrm{had}}(z)}{z}
\frac{1}{t-z} F_1(z) \nonumber \\
&+&
c \int_{4M_{\pi}^2}^{\infty} \frac{dz}{z\left(s-z\right)}
\Bigl\{
R_{\mathrm{had}}(z)\Bigl[F_2(z)
\nonumber \\ 
&+& F_3(z)\ln \bigl| 1- \frac{z}{s}\bigr| \Bigr] -
  R_{\mathrm{h}}  (s)\Bigl[F_2(s) \nonumber \\
&+& F_3(s)\ln \bigl| 1-\frac{z}{s} \bigr| \Bigr]
\Bigr\} 
 + c~\frac{R_{\mathrm{h}}  (s)}{s} \times \nonumber \\
&&\Bigl\{
F_2(s)\ln\Bigl(\frac{s}{4M_{\pi}^2}-1\Bigr)
- 6\zeta_2 F_{a}(s) \nonumber \\
&+&F_3(s)\Bigl[
2\zeta_2
+\frac{1}{2}\ln^2\Bigl(\frac{s}{4M_{\pi}^2}-1\Bigr) \nonumber \\
&+&\text{Li}_2\Bigl(1-\frac{s}{4M_{\pi}^2}\Bigr)
\Bigr]
\Bigr\} , 
\end{eqnarray}
with $c=\alpha^4/(\pi^2 s)$.
The $F_{1,2,3}$ and $F_a$ are defined as in \cite{Actis:2007fs} and \cite{webPage:2007xx}, and
$R_{\mathrm{h}}(s) = \theta(s-4M_{\pi}^2)~R_{\mathrm{had}}(s)$.

Numerical results are given in Tables~\ref{nums1} and \ref{nums2},
where we include also the QED Born prediction, the Standard Model effective Born 
prediction for $M_Z= 91.188 \mathrm{~ GeV}$, $\Gamma_Z=2.495  \mathrm{~ GeV}$, $\sin^{2,eff}_W=0.23$  and the contribution from the running of the fine-structure
constant. The cut on the energy of the soft photons is set to $\sqrt{s}\slash 2$.
We can see that, refering to the per mil accuracy:  (i) electron vertices
dominate over the rest of vertices (however, it is known that they largely cancel with the
contribution of the soft electron pair emission of \cite{Arbuzov:1995vj},
see also  \cite{hadrPRD}); (ii) contributions from infrared-safe boxes in Eq.~\ref{sig1} are 
substantial, mostly due to the factorizing diagrams;
(iii) hadronic contributions play an important role.

\begin{table}[ht]
\setlength{\arraycolsep}{\tabcolsep}
\renewcommand\arraystretch{1.1}
\begin{tabular}{|r|r|r|r|r|}
\hline
$\sqrt{s}$ & 1 GeV& 10 GeV& $M_Z$  & 500 GeV\\
\hline 
\hline
     $e$ &  -45.87  &  -124.2 &   -254.4  &  -400.6 \\
\hline 
     $\mu$ & {\bf 0.36}   & {\bf -4.8}  & {\bf -29.1}   & {\bf - 70.1 }  \\
           & 0.21   &   -4.8  &    -29.1   &  -70.1  \\
\hline 
    $\tau$ &  {\bf 0.02}   & {\bf 0.3}   &  {\bf -2.1}  & {\bf -13.5}  \\
           & $\times$   & 0.1  &  -2.1   & -13.5\\
\hline 
       $t$ &  & $<$ {\bf 10}$^{-1}$  & $<$ {\bf 10}$^{-1}$    &  {\bf 0.3}  \\
           & $\times$   & $\times$  & $\times$   & $<$ {\bf 10}$^{-1}$ \\
\hline
had.     &   {\bf 0.92}      &    {\bf -4.8}      &     {\bf -57.1}    &    {\bf -165.3}     \\
\hline 
\end{tabular}
\hspace*{2mm}
\caption[]{\em Contributions to Re\,$V_2$ in the $s$ channel.
           See  Table~\ref{table:ver1} for further details.}
\label{table:ver2}
\end{table}

\section{Summary}
Virtual NNLO QED corrections to massive Bhabha scattering have been completed in the small 
electron mass limit. Photonic, electron and heavy-fermion contributions have been checked by 
independent groups and different methods of calculations. Hadronic contributions have been
calculated through the dispersion relation approach; the kernels employed have been
checked through a comparison with the heavy-fermion result of \cite{Bonciani:2008zz,Bonciani:2008ep}.

The analysis of the virtual NNLO contributions shows that the results can influence 
Bhabha physics; therefore, further studies including Monte Carlo calculations with real 
bremsstrahlung are welcome \cite{gmontagna}.

\begin{table*}[t]\centering
\setlength{\arraycolsep}{\tabcolsep}
\renewcommand\arraystretch{1.1}
\begin{tabular}{|r|r|r|r|r|}
\hline 
$\sqrt{s}$ [GeV] & 1& 10& $M_Z$ & 500\\
\hline 
\hline
{\rm QED Born} 
 & 440994  & 4409.94   & 53.0348   & 
 1.76398  \\ 
\hline 
\hline
{\rm ew. Born} 
& 440994 & 4409.95  & 53.0370  & 1.76331
\\ 
\hline
\hline
 {\rm self energies (A)} & 445283 & 4495.45 & 55.5352 & 1.90910 \\ 
\hline 
\hline
{\rm irred. vertices (B)}
& -56 & -2.74 & -0.1005 & -0.00704 \\ 
\hline
\hline 
{\rm boxes+red. (C)}\qquad $e$ &193   & 5.73  & 0.1357  &0.00673   \\
\hline 
$\mu$ & $<$ {\bf 1}  & {\bf 0.42}  & {\bf 0.0408} & {\bf 0.00288}  \\ 
 &  $- $ & 0.08   &0.0407  & 0.00288  \\
\hline 
$\tau$ &  $<$ {\bf 1}   &  $<$ {\bf 10$^{-2}$}  & {\bf 0.0027} & {\bf 0.00088} \\ 
& $-$ & $-$  &-0.0096  & 0.00084 \\
\hline 
$t$ & $<$ {\bf 1}  & $<$ {\bf 10$^{-2}$}  &   $<$ {\bf 10$^{-4}$}  & $<$ {\bf 10$^{-5}$}  \\ 
& $-$ & $-$ & $-$ & $-$  \\
\hline 
had & $<$ {\bf 1}  & {\bf 0.39} &  {\bf 0.0877} & {\bf 0.00811} \\ 
\hline 
\hline 
  $ \sum_{\rm{boxes+red.}}$ (C)& 193   & 6.54  & 0.2669 & 0.01860    \\ 
\hline 
\hline 
\hline 
 $\mathrm{ A+B+C}$ & 445420  & 4499.25 & 55.7016 & 1.92066    \\ 
\hline 
\end{tabular}
\caption[]{%
Differential cross sections
in nanobarns at a scattering angle $\theta=3^\circ$, in units of $10^2$.
A -- QED $\alpha$ running \cite{Eidelman:1995ny}; B -- irreducible vertex corrections,
C -- net sum of infrared-sensitive corrections, including double boxes,
dispersion approach (first line) and analytical expansion \cite{Actis:2007gi} (second line).
When $m_f^2> s,|t|,|u|$, the entry is suppressed.
Parameters:
$\omega= \sqrt{s}\slash 2$, $M_Z$ = 91.1876 GeV, $m_t$=172.5 GeV.%
}
\label{nums1}
\end{table*}

\begin{table*}[bht]\centering
\setlength{\arraycolsep}{\tabcolsep}
\renewcommand\arraystretch{1.1}
\begin{tabular}{|r|r|r|r|r|}
\hline 
$\sqrt{s}$ [GeV] & 1& 10& $M_Z$& 500\\
\hline 
\hline
{\rm QED Born}& 466537 & 4665.37  &56.1067  &1.86615  \\ 
\hline
\hline
{\rm ew. Born}& 466558 & 4686.27  & 1289.3011  & 0.85496  \\ 
\hline
\hline
  {\rm self energies (A)} & 480106 & 4984.83 & 62.9027 & 2.17957 \\ 
\hline 
\hline
{\rm irred. vertices (B)} & -494 & -14.35 & -0.4239 & -0.02602 \\ 
\hline
\hline
{\rm boxes + red. (C)}\quad $e$ & 807  & 14.53   & 0.2706 & 0.01193  \\
\hline 
$\mu$ & {\bf 160}    & {\bf 6.08}  & {\bf 0.1470}  &  {\bf 0.00726} \\ 
      & 153          & 6.08        & 0.1470        & 0.00726   \\
\hline 
$\tau$ & {\bf 2}   & {\bf 1.33} & {\bf 0.0752} 
& {\bf 0.00457}  \\ 
& $-$ & 1.07   & 0.0752  & 0.00457  \\
\hline 
$t$ & $<$ {\bf 1}  & $<$ {\bf 10$^{-2}$}  & {\bf 0.0005}  & {\bf 0.00043}   \\ 
& $-$ & $-$ & $-$ & -0.00013 \\
\hline 
had. & {\bf 234} 
& {\bf 16.07} 
&{\bf 0.4701} 
&
{\bf 0.02461} 
\\ 
\hline 
\hline 
 $ \sum_{\rm{boxes+red.}}$ (C) & 1203     & 38.01      & 0.9634     &  0.04880
  \\ 
\hline 
\hline 
 $\mathrm{ A+B+C}$   & 480815  & 5008.49 & 63.4422 & 2.20235   \\ 
\hline 
\hline 
\end{tabular}
\caption[]{
The same as in Tab.~\ref{nums1}, at a scattering angle $\theta=90^\circ$, in units of $10^{-4}$.
\label{nums2}
}
\end{table*}

\section*{Acknowledgements}
We would like to thank B.~Kniehl and H.~Burkhardt for help concerning
$R_{\mathrm{had}}$.
Work supported in part by Sonderforschungs\-be\-reich/Transregio 9--03 of DFG
`Computer\-ge\-st{\"u}tzte Theo\-re\-ti\-sche Teil\-chen\-phy\-sik', 
and by MRTN-CT-2006-035505 ``HEPTOOLS'' and MRTN-CT-2006-035482 ``FLAVIAnet''.

\providecommand{\href}[2]{#2}


\end{document}